\documentclass{v23windows}
\usepackage{mathtools,amsmath,empheq,slashed}

\def\be{\begin{equation}}
\def\ee{\end{equation}}
\def\bea{\begin{eqnarray}}
\def\eea{\end{eqnarray}}

\newcommand{\bib}[1]{Ref.~\cite{#1}}

\newcommand{\fig}[1]{Fig.~\ref{#1}}

\begin{document}
\vspace*{4cm}
\title{Toward global fits using Higgs STXS data with Lilith}

\author{Dang Bao Nhi Nguyen$^{a,b}$, \underline{Duc Ninh Le}$^{c}$, Sabine Kraml$^d$, Quang Loc Tran$^{a,b}$, Van Dung Le$^{a,b}$}


\address{
$^{a}$Department of Theoretical Physics, University of Science, Ho Chi Minh City 70000, Vietnam\\
$^{b}$Vietnam National University, Ho Chi Minh City 70000, Vietnam\\
$^{c}$Faculty of Fundamental Sciences, PHENIKAA University, Hanoi 12116, Vietnam\\
$^{d}$Laboratoire de Physique Subatomique et de Cosmologie, Universit\'e Grenoble-Alpes,\\ CNRS/IN2P3, 53 Avenue des Martyrs, F-38026 Grenoble, France
}

\maketitle\abstracts{In this talk, we present the program Lilith, a python package for constraining 
new physics from Higgs measurements. We discuss the usage of signal strength results in the latest 
published version of Lilith, which allows for constraining deviations from SM Higgs couplings through 
coupling modifiers.  Moreover, we discuss the on-going development to include Higgs STXS data and 
SMEFT parametrizations in Lilith with the aim of performing global fits of the ATLAS and CMS data. 
As we point out, detailed information on Standard Model uncertainties and their correlations 
is important to enable the proper reuse of the experimental results.
}

\section{Introduction}
The discovery of the Higgs boson by ATLAS and CMS in 2012 at CERN has opened a new window to the universe, 
in particular for the search for physics beyond the \mbox{Standard Model (SM)}.  Of particular interest are 
the couplings of the 125~GeV  Higgs boson to heavy fermions (top, bottom quarks) and to 
massive gauge bosons ($W$ and $Z$). The current ATLAS and CMS results from Run~2 of the LHC already 
provide very stringent tests of these couplings. 
The data types concerning us in this contribution are Higgs signal strengths (SS) and Simplified Template Cross Sections (STXS), 
which are published in many Higgs-analysis papers by ATLAS and CMS since the Higgs discovery; 
see, e.g., \bib{yellow4} for a detailed discussion.

The SS are defined as
\begin{align}
\mu_i^j=\dfrac{(\sigma_i\times\mathcal{B}^j)_\text{exp}}{(\sigma_i\times\mathcal{B}^j)_\text{SM}},
\end{align}
where $\sigma_i$ is the inclusive Higgs production cross section of the channel $i$, which can be 
$ggF$ ($87\%$), $VBF$ ($7\%$), $VH$ ($4\%$), $t\bar{t}H$ ($1\%$), $tH$ ($<0.1\%$). These cross sections are usually 
measured with $|y_H|<2.5$. $\mathcal{B}^j$ is the Higgs branching fraction in the decay mode $j$, which can be 
$b\bar{b}$, $WW^*$, $ZZ^*$, $\gamma\gamma$, etc. 
The best-fit values for $\mu_i^j$ together with their uncertainties and correlations, as reported in ATLAS and CMS publications, 
can be used to constrain deviations of the Higgs couplings from their SM expectations. 

The SS data type is simple and has been used efficiently in many experimental and theoretical papers to check the consistency 
of the SM with the ATLAS and CMS data. 
Its power for constraining new physics is, however, limited for (essentially) two reasons:  
1) the dominant contribution to the inclusive cross section comes from the low energy region, while new physics is expected to affect the high energy region, e.g., when $p_T^H$ is large; and 2) the SS framework explicitly assumes that the signal acceptances are the same as in the SM. 
The latter means that the SS framework is valid only for overall scalings, but not for changes in kinematic distributions. 
The STXS scheme was introduced to overcome these limitations without going all the way to fiducial differential distributions.  
Concretely, the STXS split the inclusive phase space into different regions, called \emph{bins}, corresponding to different ranges of $p_T^H$, jet multiplicity, etc. 
Similar to the SS results, ATLAS and CMS have been providing best-fit values together with uncertainties and correlations for the STXS.   

\section{Signal strength data and reduced couplings in Lilith}

Lilith, {\bf Li}ght {\bf Li}kelihood Fi{\bf t} for the {\bf H}iggs, is a python package for constraining new physics 
from Higgs measurements~\cite{Bernon:2015hsa,Kraml:2019sis}. 
The current stable release Lilith-2.1 includes a database with the complete ATLAS and CMS SS data from Run~1 and
from Run 2 with 36/fb, plus some old Tevatron results. 
The core of the program is the likelihood calculation based on multi-dimensional Gaussian distributions 
of variable width to deal with correlated, asymmetric uncertainties~\cite{Kraml:2019sis}. 
A global likelihood is built from this by trivially combining the individual experimental results in the database. 

\begin{figure}[h!]\centering
  \includegraphics[width=0.3\textwidth]{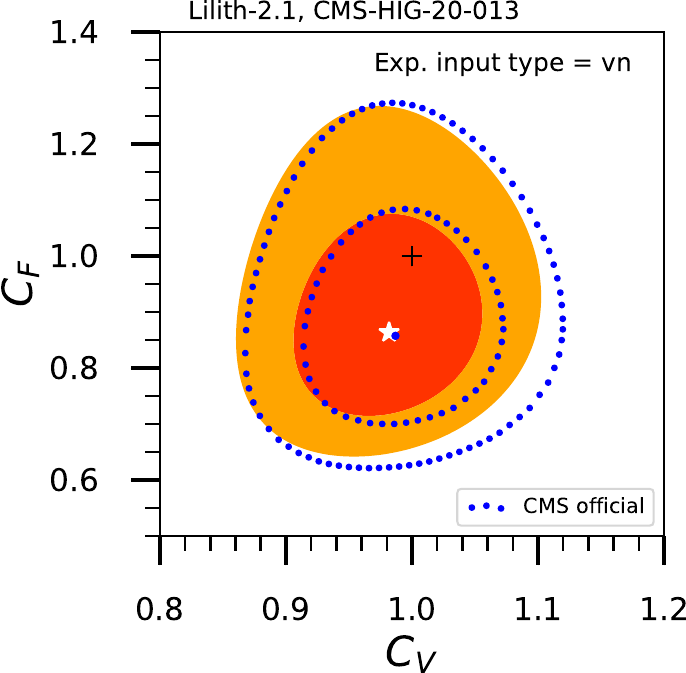}\quad 
  \includegraphics[width=0.31\textwidth]{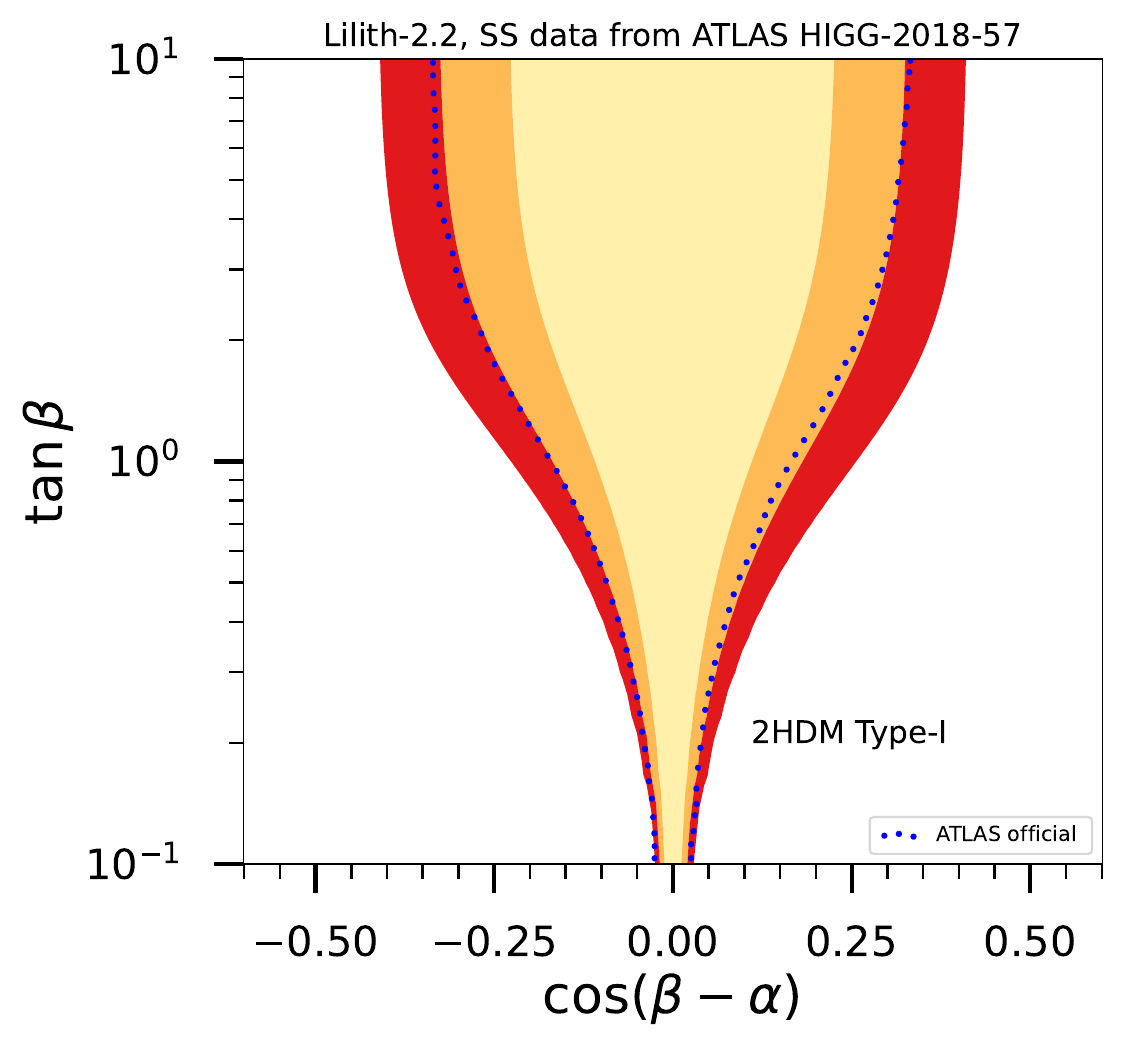}
  \caption{Comparisons of Lilith results (colored regions) using SS data against ATLAS and CMS results (dotted contours). 
   On the left, the red and orange regions are $68\%$ and $95\%$ CL regions, respectively, from Lilith, while the dotted contours are 
   the corresponding results from CMS-HIG-20-013. 
   On the right, the light yellow, orange and red regions correspond to $68\%$, $95\%$ and $99.7\%$ CL, respectively, from Lilith, 
   while the dotted contour is the $95\%$ CL result from ATLAS-HIGG-2018-57.}
  \label{fig_ss_validation}
\end{figure}

Within Lilith, the SS are expressed in terms of coupling modifiers $C_X$ ($X=t,b,c,W,Z,....$), 
which rescale the Higgs couplings to SM particles with respect to their SM values~\cite{Bernon:2015hsa}. 
Likelihood fits can thus be performed in terms of SS or in terms of the reduced couplings $C_X$. 
An example for the latter is shown in \fig{fig_ss_validation} (left), which compares the Lilith fit of universal 
scaling factors for the couplings to fermions, $C_F$, and the coupling to massive gauge bosons, $C_V$, 
to the corresponding results from the CMS measurement in the $H\to WW^*\to 2\ell 2\nu$ decay channel~\cite{CMS:2022uhn}. 
We note that such comparisons are particularly useful for validations of the database and of our 
statistical method.  

This approach can be used for constraining the parameters of any new physics model that has the 
same tensor structure as the SM, such that the couplings of the SM-like Higgs simply rescale. 
The simplest case are two Higgs doublet models (2HDMs), where, for decoupled $H^\pm$, 
the Higgs couplings depend on only on the two angles $\alpha$ and $\beta$.   
As an example, \fig{fig_ss_validation} (right) shows constraints in the $\tan\beta$ vs.\ $\cos(\beta-\alpha)$ plane for the 
2HDM of Type~I, where $C_V=\cos(\beta-\alpha)$ and $C_F=\cos\alpha/\sin\beta$, comparing to the official results from ATLAS \cite{ATLAS:2019nkf}. 

\section{Extension of Lilith to STXS data and SMEFT fits}

\begin{figure}[h!]\centering
   \includegraphics[width=0.305\textwidth]{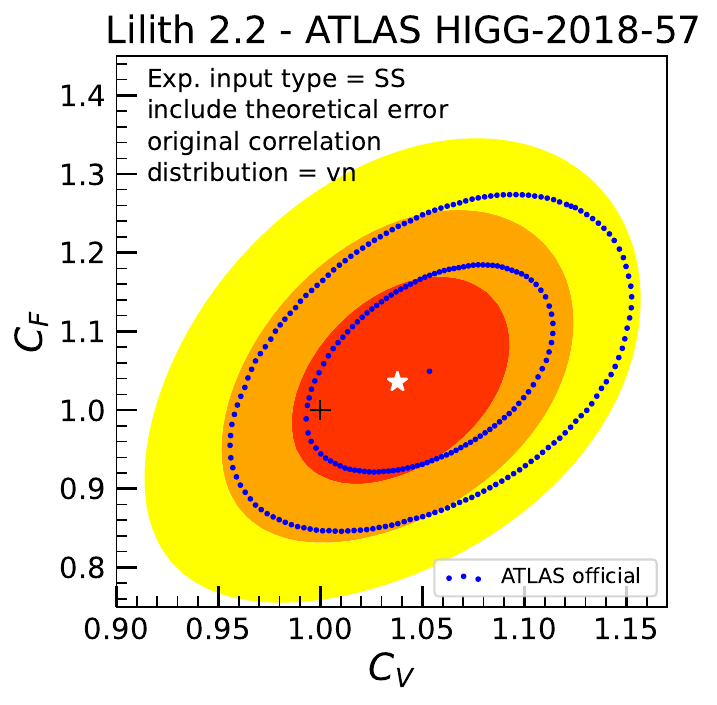}
   \includegraphics[width=0.3\textwidth]{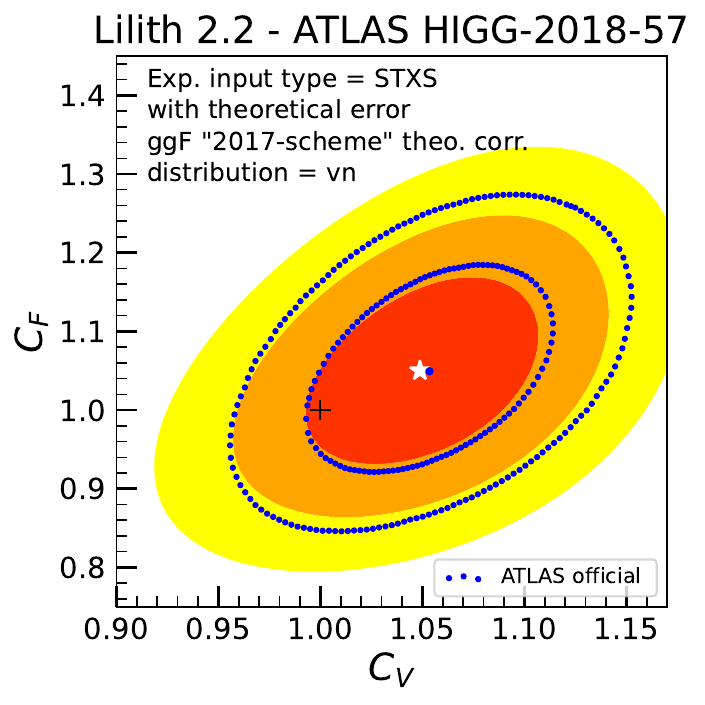}
   \caption{Comparison of a fit of $C_F$ vs.\ $C_V$ using SS (left) and STXS (right) data from the 
   ATLAS combined measurements of Higgs boson production and decay with up to 80/fb, ATLAS-HIGG-2018-57.
   The red, orange and yellow regions are the $68\%$, $95\%$ and $99.7\%$ CL regions from Lilith, 
   while the blue dotted contours are the $68\%$ and $95\%$ CL results from ATLAS.}
   \label{fig_stxs_vs_ss}
\end{figure}

We aim to go beyond the SS database and include important new physics parametrizations. 
The step taken in this work is to implement the STXS data type and SM effective field theory (SMEFT) parametrizations in Lilith. 
A first comparison between the SS and STXS data types is shown in \fig{fig_stxs_vs_ss}, still using the reduced coupling scheme. 
We can see clearly that the STXS data allow to better reproduce the official ATLAS result.  

\begin{figure}[h!]\centering
\includegraphics[width=0.3\linewidth]{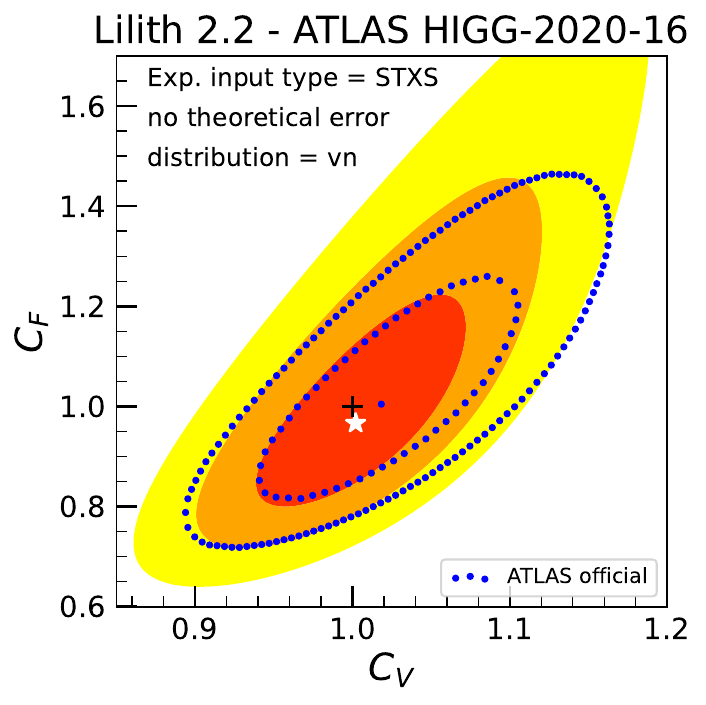}
\includegraphics[width=0.3\linewidth]{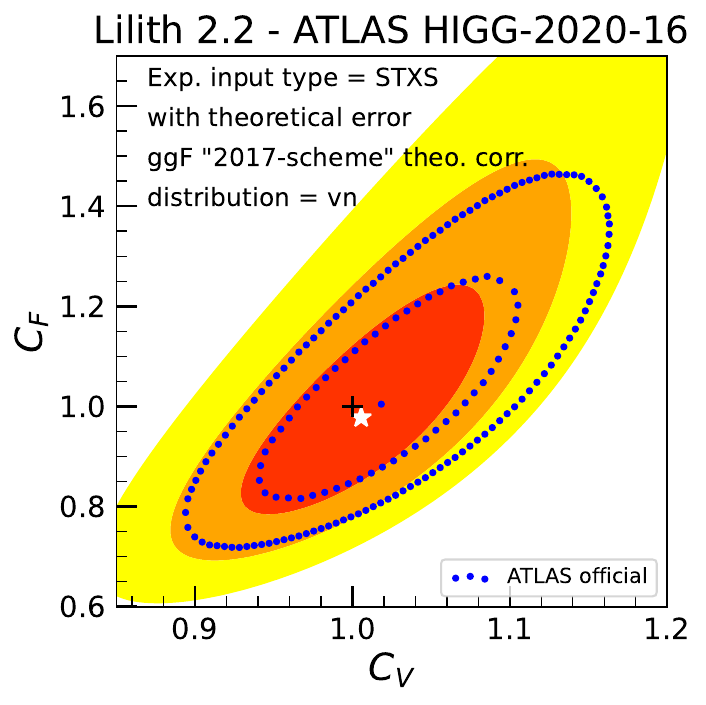}
\includegraphics[width=0.3\linewidth]{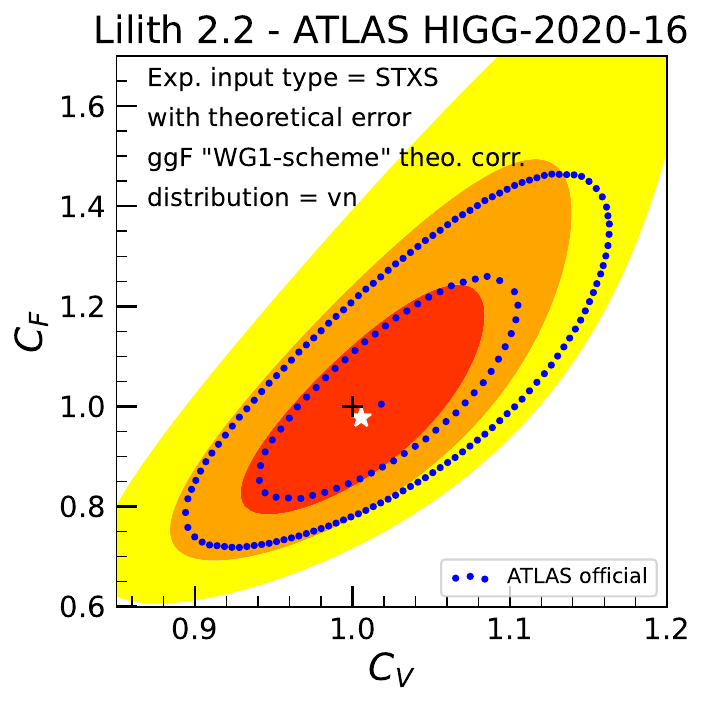}  
\caption{Comparisons between different treatments of theoretical uncertainties, for the STXS data of ATLAS-HIGG-2020-16. 
Left: without SM errors, middle: SM errors included assuming the ``2017-scheme", right: 
SM errors included assuming the ``WG1-scheme". The results for the ``JVE" and ``STXS" schemes are not shown but almost identical to those of the middle and right plots.}
  \label{fig_theory_corr}
\end{figure}

While the STXS result in \fig{fig_stxs_vs_ss} is encouraging, there is still a sizeable discrepancy between 
Lilith and ATLAS. This discrepancy  is even more pronounced for the ATLAS measurement 
in the $H\to\gamma\gamma$ channel with 139/fb, ATLAS-HIGG-2020-16~\cite{ATLAS:2022tnm}, shown in \fig{fig_theory_corr}.  
We think that the discrepancy comes to large extent from the treatment of the theoretical uncertainties, 
and in particular from unknown theory correlations.  
The problem here is that the determination of the theory correlations 
is not unique, as it depends on the uncertainty sources and assumptions on 
their correlations (see Section {\bf I.4.2.a} of \bib{yellow4} for a nice discussion on this issue). 
Indeed there exist various approximations of theory correlations without a guidance which one is the best choice. 
For the $ggF$ bins, these approximations are called ``2017-scheme", ``JVE-scheme", ``STXS-scheme", ``WG1-scheme", 
provided in \bib{ggF_unc_schemes} 
(see also the LHC HXSWG webpage~\cite{LHCHXSWG_stxs}). 
For the ATLAS-HIGG-2020-16 data, we observe that the above 
approximations give almost identical results, see \fig{fig_theory_corr}, without completely resolving the discrepancy with the ATLAS fit.

The way forward, in our opinion, would be that the SM uncertainties \emph{and} at least one approximation for their correlations 
are provided together with the STXS measurements. As of now, ATLAS and CMS have published the best-fit values together 
with measurement uncertainties and their correlations. Sometimes, the SM predictions and their uncertainties are 
also provided, but information on the assumed shapes and correlations are missing. Without this information, the usage 
of the STXS measurements to constrain new physics is severely limited.     

\begin{figure}[h!]\centering
  \includegraphics[width=0.43\textwidth]{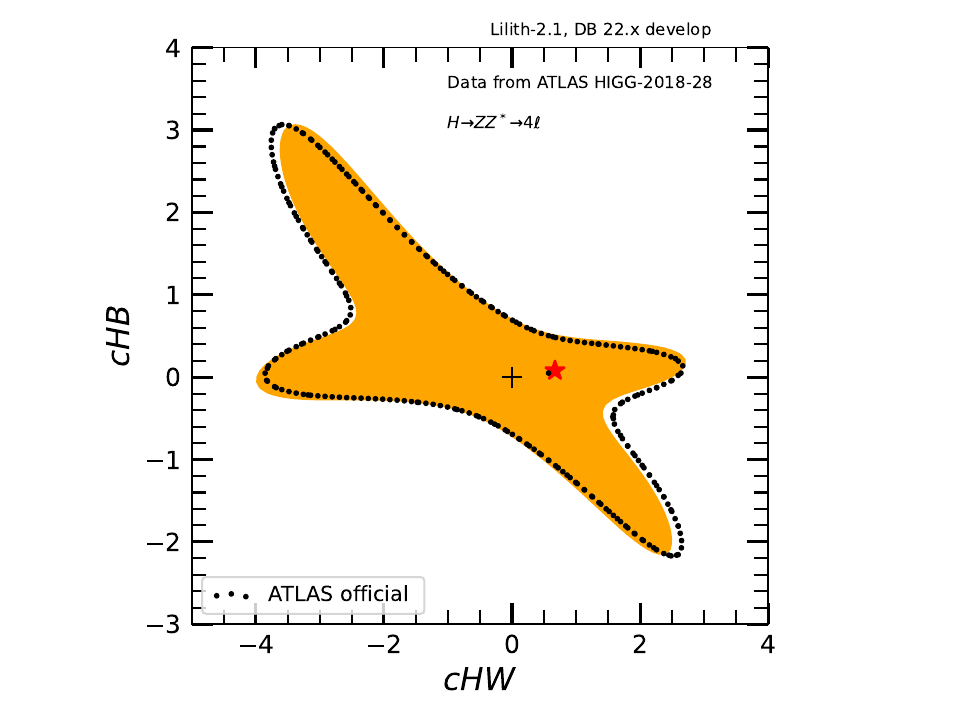}
  \includegraphics[width=0.43\textwidth]{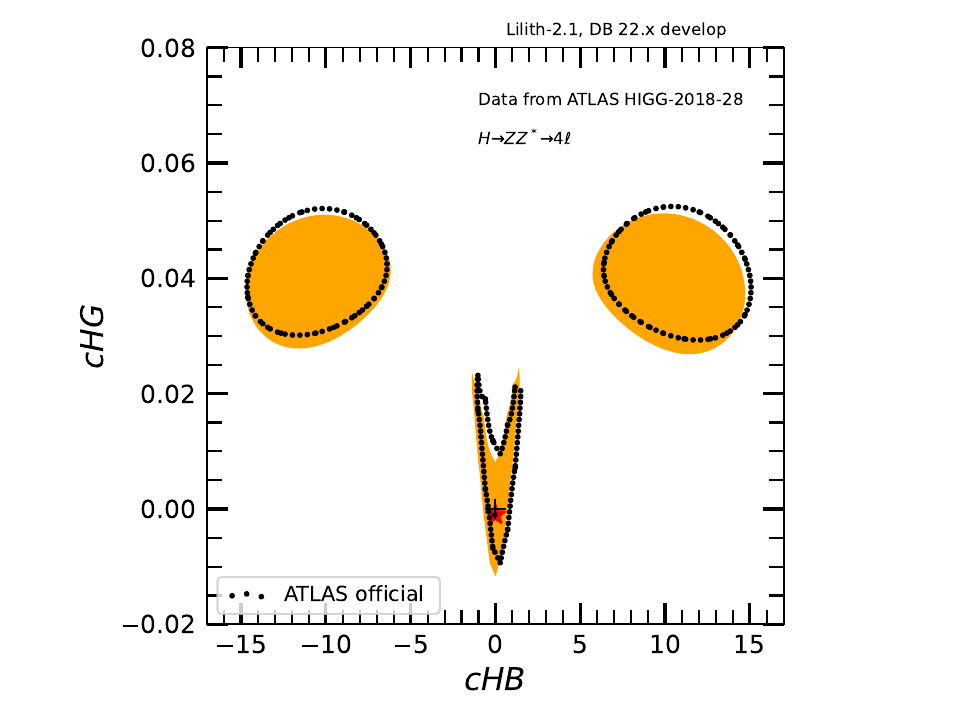}
  \caption{Comparison between Lilith (using the ``2017-scheme" $ggF$ SM correlations) and ATLAS for a SMEFT parametrization. Details of this parametrization and of the Wilson coefficients $cHB$, $cHW$, $cHG$ are provided in the ATLAS-HIGG-2018-28 paper.}
  \label{fig_smeft}
\end{figure}

Another ongoing development in Lilith is the extension to SMEFT fits. To this end, the parametrization of the STXS bins 
in terms of Wilson coefficients of the SMEFT is necessary. This is available from the ATLAS measurement in the $H\to ZZ^*\to 4\ell$ 
channel with 139/fb, ATLAS-HIGG-2018-28~\cite{ATLAS:2020rej}. In \fig{fig_smeft}, we show the SMEFT fit results obtained with Lilith using the data from ATLAS-HIGG-2018-28; as can be seen, they agree very well with the official ATLAS results. 

\section{Conclusions}
We have discussed the usage of Higgs SS and STXS data for constraining new physics from Higgs measurements, 
as well as the program Lilith, which has been developed for this purpose. 
The current stable release of Lilith makes use of SS data, and likelihood fits are based on rescaling of the Higgs couplings with respect to the SM. 
Ongoing developments include the 
extension of Lilith with STXS data and likelihood fits within the SMEFT, for both of which first results have been presented here.  
We have also shown that more information on SM uncertainties and their correlations 
would be important for the proper reuse of the experimental results.

\section*{Acknowledgement}
This research is funded by the  Vietnam National Foundation for Science and Technology Development (NAFOSTED) 
under grant number 103.01-2020.17. SK\ acknowledges funding by the IN2P3 theory master project BSMGA and the 
ANR project EFTatLHC, ANR-22-CE31-0022-02.  

\section*{References}
\bibliographystyle{unsrt}

\begin{thebibliography}{99}

\bibitem{yellow4}
D.~de Florian \textit{et al.} [LHC Higgs Cross Section Working Group],
\emph{{Handbook of LHC Higgs Cross Sections: 4. Deciphering the Nature of the Higgs Sector}},
[\href{https://arxiv.org/abs/1610.07922}{{\ttfamily 1610.07922}}].

\bibitem{Bernon:2015hsa}
J.~Bernon and B.~Dumont, \emph{{Lilith: a tool for constraining new physics
  from Higgs measurements}},
  \href{https://doi.org/10.1140/epjc/s10052-015-3645-9}{\emph{Eur. Phys. J.}
  {\bfseries C75} (2015) 440}
  [\href{https://arxiv.org/abs/1502.04138}{{\ttfamily 1502.04138}}].

\bibitem{Kraml:2019sis}
S.~Kraml, T.~Q.~Loc, D.~T.~Nhung and L.~D.~Ninh,
\emph{{Constraining new physics from Higgs measurements with Lilith: update to LHC Run 2 results}},
\emph{SciPost Phys.} \textbf{7} (2019) no.4, 052
[\href{https://arxiv.org/abs/1908.03952}{{\ttfamily 1908.03952}}].

\bibitem{CMS:2022uhn}
A.~Tumasyan \textit{et al.} [CMS],
\emph{{Measurements of the Higgs boson production cross section and couplings in the W boson pair decay channel in proton-proton collisions at $\sqrt{s}=13\,\text {Te\hspace{-.08em}V}$}},
\emph{Eur. Phys. J. C} \textbf{83} (2023) no.7, 667, report CMS-HIG-20-013,
[\href{https://arxiv.org/abs/2206.09466}{{\ttfamily 2206.09466}}].

\bibitem{ATLAS:2019nkf}
G.~Aad \textit{et al.} [ATLAS],
\emph{{Combined measurements of Higgs boson production and decay using up to $80$ fb$^{-1}$ of proton-proton collision data at $\sqrt{s}=$ 13 TeV collected with the ATLAS experiment}},
\emph{Phys. Rev. D} \textbf{101} (2020) no.1, 012002, report ATLAS-HIGG-2018-57,
[\href{https://arxiv.org/abs/1909.02845}{{\ttfamily 1909.02845}}].

\bibitem{ATLAS:2022tnm}
G.~Aad \textit{et al.} [ATLAS],
\emph{{Measurement of the properties of Higgs boson production at $\sqrt{s} = 13$ TeV in the $H\to\gamma\gamma$ channel using $139$ fb$^{-1}$ of $pp$ collision data with the ATLAS experiment}},
\emph{JHEP} \textbf{07} (2023), 088, report ATLAS-HIGG-2020-16,
[\href{https://arxiv.org/abs/2207.00348}{{\ttfamily 2207.00348}}].

\bibitem{ATLAS:2020rej}
G.~Aad \textit{et al.} [ATLAS],
\emph{{Higgs boson production cross-section measurements and their EFT interpretation in the $4\ell $ decay channel at $\sqrt{s}=$13 TeV with the ATLAS detector}},
\emph{Eur. Phys. J. C} \textbf{80} (2020) no.10, 957, report ATLAS-HIGG-2018-28,
[\href{https://arxiv.org/abs/2004.03447}{{\ttfamily 2004.03447}}].

\bibitem{ggF_unc_schemes}
Dag Gillberg, 
\href{https://dgillber.web.cern.ch/dgillber/ggF_uncertainty_2017/}
{https://dgillber.web.cern.ch/dgillber/ggF\_uncertainty\_2017/}

\bibitem{LHCHXSWG_stxs}
\href{https://twiki.cern.ch/twiki/bin/view/LHCPhysics/LHCHWGFiducialAndSTXS}
{https://twiki.cern.ch/twiki/bin/view/LHCPhysics/LHCHWGFiducialAndSTXS}

\end{thebibliography}

\end{document}